\documentclass[a4paper,amsmath,amssymb,color,nofootinbib,preprint,tightenlines,fontsize=8pt]{revtex4-2}
\usepackage{graphicx}
\usepackage[dvipsnames]{xcolor}
\usepackage[mathscr]{euscript}
\usepackage{slashed}
 \usepackage{booktabs}
 \usepackage{subfigure}
\usepackage{tablefootnote}
\usepackage[normalem]{ulem}
\usepackage{cancel}
\usepackage{caption}
\usepackage{subcaption}
\usepackage{braket} 
\usepackage[colorlinks=true,pdfstartview=FitV,
citecolor=blue,linkcolor=purple,urlcolor=Sepia]{hyperref}
\usepackage[capitalise]{cleveref}
\usepackage{orcidlink}
\usepackage{xcolor}

\newcommand{\N}{\mathcal{N}}
\begin{document}
	\baselineskip=17pt \parskip=3pt
	
	\newcommand{\XGH}[1]{{\color{red}XGH: #1}}
	\newcommand{\XDM}[1]{{\color{orange}XDM: #1}}
	\newcommand{\GV}[1]{{\color{blue}GV: #1}}
	\newcommand{\JT}[1]{{\color{purple}JT: #1}}
	\begin{center}
		\textbf{\large  Hadrophilic inelastic freeze-in dark matter in $q_1-q_2$ gauge extension and the high energy LZ event} 
	\end{center}
	
	\author{Xiao-Gang He\,\orcidlink{0000-0001-7059-6311}$,^{1,2,}$\footnote{\email{hexg@sjtu.edu.cn}}
    Xuan Hong\,\orcidlink{0009-0009-8123-6288}$,^{1,2,}$\footnote{\email{hxxh123@sjtu.edu.cn}~(corresponding author)}  
		and Sk Jeesun\,\orcidlink{0009-0005-2344-9286}$,^{1,2,}$\footnote{\email{jeesun@sjtu.edu.cn}~(corresponding author)
		} 
		\vspace{1ex} \\ \it
		$^1$State Key Laboratory of Dark Matter Physics, Tsung-Dao Lee Institute \& School of Physics and Astronomy,
		Shanghai Jiao Tong University, Shanghai 201210, China \vspace{1ex} \\
		$^2$Key Laboratory for Particle Astrophysics and Cosmology (MOE) \& Shanghai Key Laboratory for Particle Physics and Cosmology,
		Tsung-Dao Lee Institute \& School of Physics and Astronomy, Shanghai Jiao Tong University, Shanghai 201210, China \vspace{1ex} \\
		\rm Abstract 
		\vspace{8pt} \\
		\begin{minipage}{0.99\textwidth} \baselineskip=17pt \parindent=3ex \small
The recent observation of a single event at nuclear recoil energy $248\pm23\pm23$ keV by the LZ collaboration has prompted dark matter (DM) model builders to explore new physics explanations.
While most of the existing literature invokes an endothermic scattering by a TeV-scale DM  to explain the event, the favored DM interaction strength is in tension with the indirect detection constraints for most of the models.
To evade this, this work explores exothermic DM scattering in a hadrophilic $q_1-q_2$ gauge extension setup with a new $Z'$ boson as a mediator to the dark sector.
This can naturally  escape the indirect searches in the absence of tree-level coupling of the mediator to neutrinos, $W,\tau$, and $b$ quark.
We consider freeze-in production at a low reheating temperature to populate DM in the early universe, with a comparatively smaller coupling than the thermal freeze-out scenario.
We identify the $Z'$ parameter space
that can explain the observed relic density and LZ  event simultaneously.
We also discuss the relevant constraints coming from flavor mixing and FCNC.
\end{minipage}}
	\maketitle
	
	\newpage

Numerous cosmological and astrophysical observations have strongly suggested the presence of dark matter (DM) in our universe~\cite{Zwicky:1933gu,Rubin:1970zza,Clowe:2006eq,Planck:2018vyg}.
DM constitutes $25\%$ of the energy budget of the universe and plays a crucial role in structure formation \cite{Planck:2018vyg}.
Despite these multiple indications and efforts spanning several decades, the elementary nature of DM and its non-gravitational interaction with standard model (SM) particles are unknown.
To uncover the DM nucleon (electron) interactions, several ton-scale direct detection experiments such as XENONnT, LUX-ZEPLIN (LZ), and PandaX-4T are being operated worldwide~\cite{XENON:2023cxc,LZ:2024zvo,PandaX-4T:2021bab} .
Recently, LZ has observed a single event at nuclear recoil energy ($E_R$) $248\pm23\pm23$ keV, which is too high for any known SM background \cite{LZ:2026axp}.
This has motivated the community to look for possible beyond standard model (BSM) explanations. 
Obviously, the weakly interacting massive particle (WIMP) DM is one of the most well-motivated candidates to produce such isolated events.
However, the challenge for the vanilla WIMP paradigm is the associated huge recoil events at low recoil energies ($\lesssim 100$ keV) in addition to the high event at $\sim250$keV $E_R$, which will be in tension with the null observations.

To circumvent this issue, the most economical BSM framework proposed is inelastic DM~\cite{DiMauro:2026ldr,Visinelli:2026kgt,Smirnov:2026aqk,Dent:2026bji,deLima:2026shq,Yang:2026wpb,Okada:2026eol,Ahmed:2026qjg,Du:2026lpa,Borah:2026zwf,Yuan:2026djt,Zhu:2026dag,Lee:2026xxh,Lee:2026jxl,Langhoff:2026ujr,Chatterjee:2026scv,Fan:2026hzw,Qi:2026vyp,Kumar:2026lgi, Lee:2026wof,DuWang:2026LZ,RoddEtAl:2026LZ,McCabe:2026LZ,Unwin:2026LZ,GuEtAl:2026LZ,BaerBarger:2026LZ,LeeHyunMin:2026LZ,WangXiao:2026LZ,KotlarskiEtAl:2026LZ,LiangEtAl:2026LZ,DiMauroShaikh:2026LZ,DasEtAl:2026LZ,AlhazmiEtAl:2026LZ,BisalEtAl:2026LZ,AsadiEtAl:2026LZ,KhanEtAl:2026LZ,He:2026LZ,FrolovskyKetov:2026LZ,BandyopadhyayEtAl:2026LZ}.
Other possible explanations also include 
neutrino up-scattering\cite{Jeesun:2026vzo}, boosted DM \cite{KannikeEtAl:2026LZ,HeikinheimoZimmermann:2026LZ}, PBH evaporation \cite{Chattaraj:2026fxn}.
In the inelastic DM framework, a lighter DM ($\chi_1$) scatters with nucleus ($\mathcal{N}$) to produce a heavier state $\chi_2$ i.e. $\chi_1+\N\to \chi_2+\N$.
The tiny mass splitting $\delta m$ between these two DM states is small enough to produce recoil, yet large enough to suppress events at low $E_R$. 
However, the typical interaction strength and mass of DM ($\sim$ TeV) currently being proposed to produce such isolated events might be in tension with other observational constraints \cite{Pospelov:2026ewn}.
By far, the most discussed limit comes from the solar capture and subsequent annihilation to visible final states (e.g $W^+W^-, \tau \bar{\tau}, b\bar{b}, \nu \bar{\nu}$) \cite{Pospelov:2026ewn,Bose:2026ndd, Nguyen:2026lui}.
This motivates the careful modeling of the microscopic DM nature as well as careful analysis of early universe cosmology to explain the DM genesis.
Needless to say, such a scenario is hard to realize in leptophilic DM models, as $\nu$s are part of the lepton doublets.
A similar argument also holds for models featuring a  DM mediator with universal coupling to SM fermions.
In this paper, we propose a hadrophilic flavor-dependent $U(1)_X$ extension that couples the inelastic DM with the SM and consider down-scattering of the heavier DM component to produce the event.
This naturally makes the required direct detection cross-section smaller for exothermic scattering \cite{Fan:2026hzw}.
For the early universe production, we 
explore the freeze-in scenario at a reheating temperature smaller than DM mass.
In such a scenario, the typical interaction strength of the DM is stronger (weaker) than in the vanilla freeze-in (freeze-out) scenario \cite{Cosme:2023xpa}.
Moreover, this scenario helps realize a stable heavy DM component surviving to today, which is difficult in the thermal paradigm.


To accommodate the DM, we extend the SM with an abelian $U(1)_{X}$ gauge symmetry.
Among the SM fermions, only the first two generations of quarks are
charged under this symmetry, which we refer to as $q_1-q_2$ symmetry.
The particle content and the $U(1)_X, ~(X=u-c)$ gauge charges are shown in Table.\ref{tab:charge_assignments}.
We extend the SM sector with a SM singlet fermion DM $\chi$ and a singlet scalar $S$ to generate the mass of the new gauge boson $Z'$, both of which are charged under the new $U(1)_X$ symmetry.
The Lagrangian in such a setup reads as,
    \begin{equation}
        \begin{aligned}
            \mathcal{L}\supset&-\frac{1}{4}Z^{'\mu\nu}Z^{'}_{\mu\nu}+ \bar Q_{iL}i\slashed{D}Q_{iL}+ \bar u_{iR}i\slashed{D}u_{iR}+\bar d_{iR}i\slashed{D}d_{iR}\\
            & + \mathcal{L}_H + \mathcal{L}_{Yukawa} + \mathcal{L}_{DM}
        \end{aligned}
    \end{equation}
 For the dark sector we consider a fermion DM $\chi$ with a nonzero $U(1)_{u-c}$ charge and the Lagrangian is 
    \begin{equation}
        \begin{aligned}
            \mathcal{L}_{DM} =  i\bar{\chi} \gamma^{\mu}D_{\mu}\chi -m_\chi \bar{\chi} \chi - y_{S\chi} \overline{\chi^c}\chi S + h.c.
        \end{aligned}.
    \end{equation}
Needless to say, the covariant derivative of $\chi$ induces the vector current with $Z'$, which is most crucial interaction to establish a connection between SM and the dark sector. 
To generate the mass splitting, we assume the scalar $S$ has a nonzero vev $\langle S \rangle=v_s/\sqrt{2}$ and the corresponding mass structure becomes,
\begin{equation}
    \mathcal{L}^{\rm mass}_\chi = \frac{1}{2}(\bar{\chi}~~ \bar{\chi^c})\begin{pmatrix}
        m_\chi & \sqrt{2} y_{s\chi} v_s \\
       \sqrt{2}y_{s\chi} v_s & m_\chi
    \end{pmatrix} 
    \begin{pmatrix}
        \chi \\
        \chi^c
    \end{pmatrix}.
\end{equation}
After diagonalization, we get the physical dark states, 
\begin{eqnarray}
    \chi_{1,2}= \frac{1}{\sqrt{2}} (\chi \mp \chi^c ),~~m_{\chi_{1,2}} = m_\chi\mp \sqrt{2} y_{s\chi} v_s,~~\delta m=2\sqrt{2} y_{s\chi} v_s.
\end{eqnarray}
And one can also derive the interaction between the dark matter and the $Z^{'}$ in physical basis as
\begin{equation}
    \mathcal{L}_{\chi Z^{'}}=\frac{1}{2}q_\chi g_{Z^{'}}\bar \chi_{1}\gamma^{\mu}\chi_2 Z^{'}_{\mu}+h.c. .
\end{equation}

    \begin{table}[!tbh]
    \centering
    \renewcommand{\arraystretch}{1.15}
    \setlength{\tabcolsep}{12pt}
    \begin{tabular}{|c|c||c|c|}
        \hline
        \multicolumn{2}{|c||}{\textbf{Quark Sector}} &
        \multicolumn{2}{c|}{\textbf{Scalar/Dark Sector}} \\
        \hline
        \textbf{Field} & \textbf{Charge} &
        \textbf{Field} & \textbf{Charge} \\
        \hline\hline

        $(u_L,\,d_L)$ & $-1$ &
        $H_0$ & $0$ \\
        
        $u_R$ & $-1$ &
        $H_1,H_2$ & $+1 , -1$ \\
        
        $d_R$ & $-1$ &
        $H_3,H_4$ & $+2 , -2$ \\
        \hline

        $(c_L,\,s_L)$ & $+1$ &
        $S$ & $-2 q_\chi$ \\
        
        $c_R$ & $+1$ &
        $\chi$ & $q_\chi$ \\
        
        $s_R$ & $+1$ &
        & \\
        \hline

        $(t_L,\,b_L)$ & $0$ &
        & \\
        
        $t_R$ & $0$ &
        & \\
        
        $b_R$ & $0$ &
        & \\
        \hline
    \end{tabular}
    \caption{Charge assignments of the fields for the hadrophilic $q_1-q_2$ model. We follow the same strategy  for gauged $L_i-L_j$ model building in refs. \cite{He:1990pn,He:1991qd}. $S$ is a singlet scalar and the other scalars ($H_i$) are $SU(2)_L$ doublets. $\chi$ is the singlet fermionic  DM field.}
    \label{tab:charge_assignments}
\end{table}

Note that the gauge extension is flavor-dependent in the quark sector. 
With only one Higgs, the model can not generate the correct CKM matrix. Since the three quark generations carry $(-1,+1,0)$ $U(1)_X$ charge, the charge differences relevant for the Yukawa couplings are $0,\pm1,\pm2$. Therefore, we introduce additional four Higgs doublets to generate the full Yukawa interaction and the correct CKM matrix, as shown in Table-\ref{tab:charge_assignments}. For the Higgs sector, we have the following Lagrangian,
    \begin{equation}
        \begin{aligned}
            \mathcal{L}_H = \sum_{i}(D_{\mu}H_i)^{\dagger}(D^{\mu}H_i) + (D_{\mu}S)^{\dagger}(D_{\mu}S) - V(H_i , S)
        \end{aligned}
    \end{equation}
where $V(H_i,S)$ is the general higgs potential which satisfies the $U(1)_X$ gauge symmetry and include the $S,H_i$ mixing term such as  $\lambda_{SH_i}|S|^2 H_i^{\dagger}H_i$. We assume this mixing is sufficiently small to prevent the dark matter from interacting with the SM particles through Higgs portal. We also consider the additional Higgs particles to be much heavier than the TeV scale to evade the collider constraints arising from the scalar sector,  which can be realized by choosing the appropriate Higgs potential parameters.
Besides, we also assume that both Higgs fields have nonzero vacuum expectation values (VEVs) to generate the quark mass matrices, 
    \begin{equation}
        \langle H_i\rangle=\frac{v_i}{\sqrt{2}}, \sum_iv_i^2=v^2, v=246\, \rm GeV. 
    \end{equation}
However, due to the non-diagonal nature of $Z'$ quark interactions, 
it may induce flavor changing neutral currents (FCNC) which has stringent constraints from meson oscillations and meson decays.
Hence, one has to be careful about the CKM matrix structure and flavor-induced neutral currents, which will be discussed in a later part of this paper.
The VEV $v_s$ is also responsible for generating $Z'$ mass. $Z-Z'$ mixing proportional to $v_i v_s(i=1,2,3,4)$ will be generated and can be made small by letting $v_i\ll v_s$ to avoid the experimental constraint. We also consider $v_i~(1=1,2,3,4)\ll v_0$ to suppress the mixing of BSM scalars with SM Higgs.
        
As indicated previously, to avoid stringent indirect detection constraints, one way out is exothermic scattering.
However, for any inelastic DM model, the key challenge is to make the heavier component survive till the present day.
For the DM model considered here, thermal production will immediately deplete the heavier component ($\chi_2\to \chi_1+Z'^*$), leaving the lighter component to be cosmologically stable. 
On the other hand, freeze-in production can in principle populate both $\chi_2,\chi_1$ with equal abundance, each satisfying $50\%$ of the total abundance, thanks to the small 
$\chi_2-\chi_1-Z'$ coupling required for non-thermal production \cite{Heeba:2023bik}.
In such a scenario, the trade-off is the small direct detection cross-section, which fails to generate such a recoil event in LZ. 
In the intermediate region of required couplings between thermal and non-thermal DM, the DM relic still can be generated if the reheating temperature ($T_{\rm RH}$) of the universe is smaller than the DM mass \cite{Cosme:2023xpa,Silva-Malpartida:2023yks,Koutroulis:2023fgp,Barman:2024tjt,Khan:2025keb,Arcadi:2024wwg,Boddy:2024vgt,Arcadi:2024obp,Belanger:2024yoj,Arias:2025tvd,Bernal:2026clv,Ghosh:2026mda,Jahedi:2026mxa}.
 Note that currently, we have precise information only about the lower limit of $T_{RH}\gtrsim$ a few MeV \cite{Kawasaki:2000en,Ichikawa:2005vw}.
In the limit 
$T_{\rm RH}\ll m_{X}$ the Bessel function in the interaction rate becomes $
K_1\left({\sqrt{s}}/{T}\right) \propto \sqrt{{ T}/{s^{1/2}}} e^{-\sqrt{s}/T}$.
Thus, for a higher ratio of $m_\chi/T_{\rm RH}$, the interaction rate and consequently DM become exponentially suppressed.
To compensate for this one requires a larger coupling than the usual freeze-in scenario, which is interesting from the perspective of the LZ event.  
For a DM species $\chi_{1,2}$, the number-density equation is
\begin{equation}
\frac{dn_{\chi_i}}{dt}+3Hn_{\chi_i}
=
\Gamma_{q\bar q\to\chi_1\bar\chi_2}
\left[
1-\frac{n_{\chi_i}^2}{(n_{\chi_i}^{\rm eq})^2}
\right].
\end{equation}
The relevant interaction rate for
$q\bar q\rightarrow\chi_1\bar\chi_2$ is
\begin{equation}
\Gamma_{q\bar q\to\chi_1\bar\chi_2}
\approx
\frac{g_q^2T}{32\pi^4}
\int_{s_{4 m_\chi^2}}^\infty
ds\,\sqrt{s} (s-4 m_q^2)\,
\sigma_{q\bar q\to\chi_1\bar\chi_2}(s)
K_1\!\left(\frac{\sqrt{s}}{T}\right),
\end{equation}
where $g_q$ represents the degrees of freedom of the quark $q$.
The total DM abundance can be obtained as $n_\chi=n_{\chi_1}+n_{\chi_2}$.We signify $m_{\chi_{1,2}}\mp \delta\approx m_\chi$ which is a valid approximation for $m_{\chi_{1,2}}> T_{\rm }\gg \delta m$. 
We work in the regime where $m_\chi\gg m_q$, and in the initial state only $u,d,c,s$ quarks participate for the chosen DM model. Since, $m_\chi\gg m_{Z'}$, the interaction rate as well as DM density is mostly insensitive to $Z'$ mass. It is always convenient to solve the Boltzmann equation in terms of co-moving abundance $Y_\chi=n_\chi/s$ with $s$ being comoving entropy density.
The left side of the above equation then becomes $dY_\chi/dT(-sHT)$ and the initial point of the evolution is chosen to be $T_{\rm RH}$.
$\sigma_{q\bar q\to\chi_1\bar\chi_2}$ is the annihilation cross-section for $s$-channel $Z'$ mediated process.


To evaluate the differential event rate in the LZ detector we convolute in the incoming DM flux with the differential cross-section,
\begin{equation}
\frac{dR}{dE_R}
=
\frac{\rho_\chi}{m_\chi} \int_{v_{\rm min}}^{v_{\rm esc}}{\rm d}^3 v ~\frac{1}{v}f \left(\vec{v}\right)
\frac{d\sigma_{\chi \N}}{dE_R}\,
,
\end{equation}
 The detailed expression of the truncated Maxwell-Boltzmann velocity distribution $f(\vec{v})$ can be found in \cite{Essig:2015cda}.
We consider  the escape velocity of DM and earth's velocity relative to DM halo to be $v_{\rm esc} \simeq 544~{\rm km\,s^{-1}} and 
~v_{\oplus} \simeq 232~{\rm km\,s^{-1}}$, repectively.
For an exothermic scattering the minimum velocity required is given by,
\begin{equation}
v_{\min}(E_R)
=
\frac{1}{\sqrt{2m_{\N} E_R}}
\left(
\frac{m_{\N} E_R}{\mu_{\chi \N}}
-
\delta m
\right),
\end{equation}
where, $\mu_{\chi \N}
=
\frac{m_\chi m_\N}{m_\chi+m_\N}$ is the reduced mass, with $\N$ as the nucleus mass. From now on, we refer $m_\chi$ as the mass of the heavy incoming DM.
The DM nucleus differential cross-section reads as,
\begin{equation}
\frac{d\sigma_{\chi \N}}{dE_R}
=
\frac{m_A}{2\mu_{\chi \N}^2v^2}
A^2 \left(\sigma_{\rm SI} \frac{\mu_{\chi \N}^2}{\mu_{\chi p}^2} \right)
F^2(q),
\end{equation}
where $A$ represents the total number of  nucleons in the Xenon nucleus, respectively.
The DM-nucleon spin-independent cross-section in our scenario with universal $Z'$ coupling to $u,d$ is given by,
\begin{equation}
\sigma_{\rm SI}
=
\frac{\mu_{\chi p}^2}{\pi}
\left(
\frac{g_Z' q_\chi}{m_{Z'}^2}
\right)^2.
\end{equation}
$\mu_{\chi p}$ stands for DM-proton reduced mass.
Finally, 
$F^2(q)$ is the form factor with $q=\sqrt{2 m_\N E_R}$, that incorporates the nuclear structure.
We use the form factor evaluated from {\tt DMFormfactor-v6} \cite{Anand:2013yka}.

\begin{figure}[!tbh]
    \centering
    \includegraphics[scale=0.6]{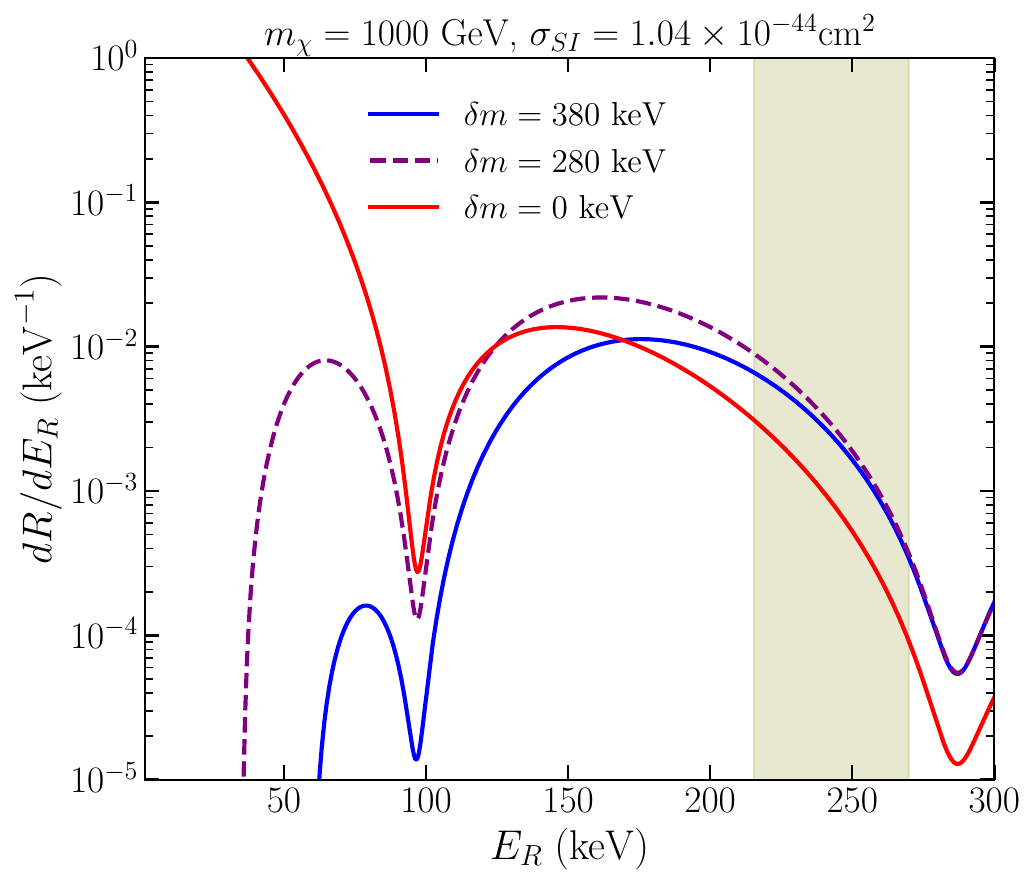}
    \caption{Expected event rate for an exothermic scattering of $m_\chi=1$ TeV,$\sigma_{SI}=1.04\times 10^{-44}~{\rm cm}^2$ and $\delta m=380,280,0$ keV shown by blue, purple and red lines. }
    \label{fig:rate}
\end{figure}
In Fig.\ref{fig:rate} we display the expected event rate for an exothermic scattering of $m_\chi=1$ TeV, $\sigma_{SI}=1.04\times 10^{-44}~{\rm cm}^2$ and $\delta m=380,280,0$ keV shown by blue, purple and red lines. Note that the second peak of the event spectra moves towards higher recoil energies with an increase in $\delta$. 
This can be easily understood from the kinematics of the inelastic  scattering.
For, $E_R\sim \delta m~ \mu_{\chi \N} E_R/m_\N$, the minimum velocity becomes 
$v_{\rm min}\to 0$, implying that a DM even at rest produces an event with the given non-zero $E_R \propto \delta m$.
\begin{figure}[!tbh]
    \centering
    \subfigure[]{\includegraphics[scale=0.45]{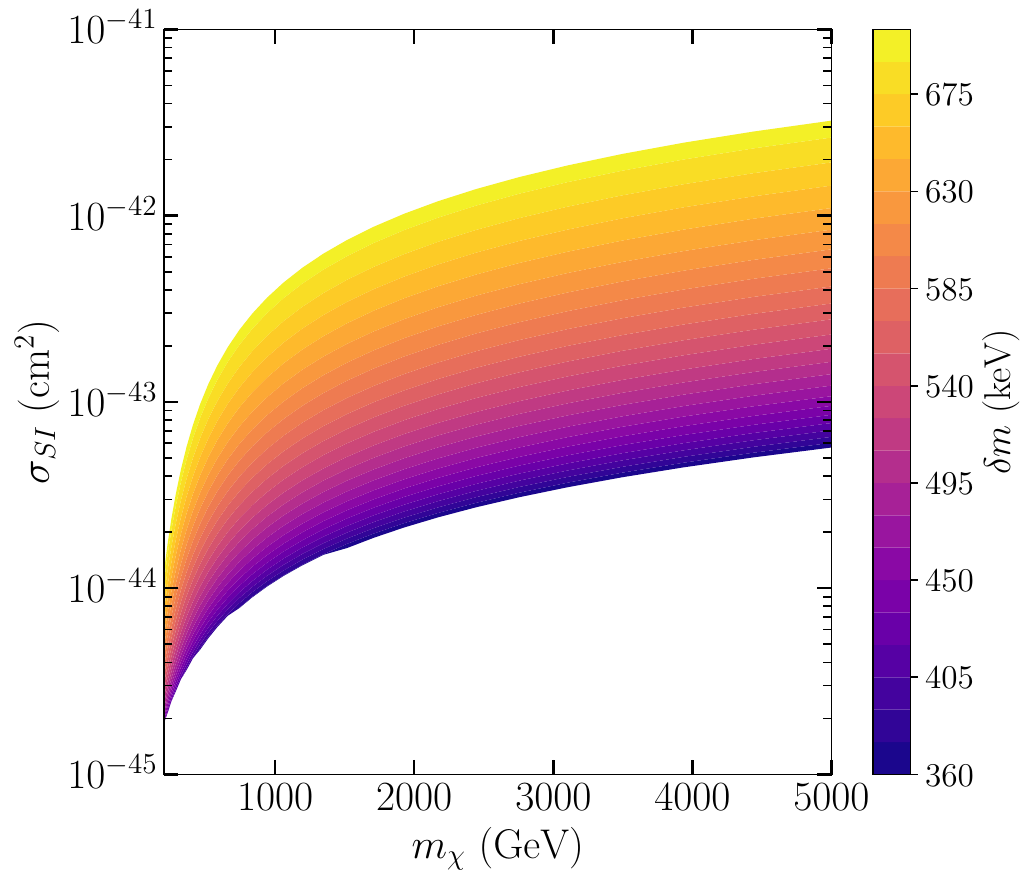}}
    \subfigure[]{\includegraphics[scale=0.45]{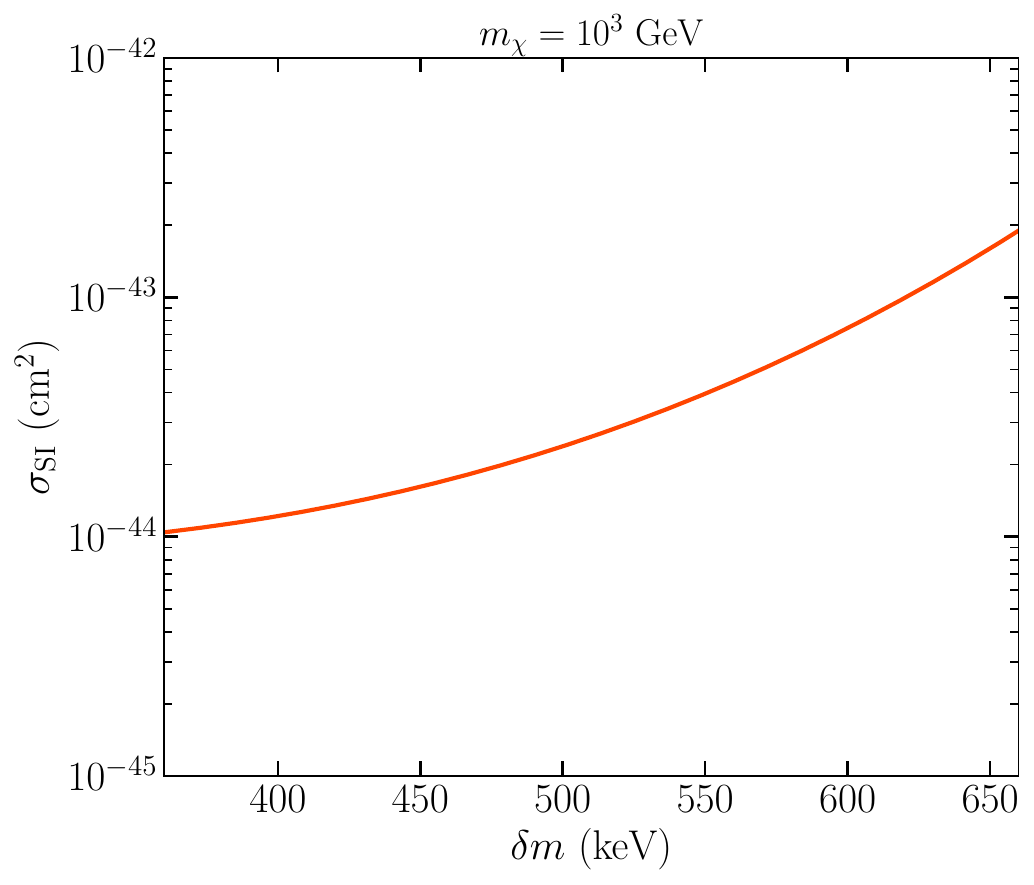}}
    \caption{(a) Required vector direct detection cross-section with varying DM mass for exothermic scattering. The color map indicates the variation in $\delta m$. (b) Required vector direct detection cross-section with varying $\delta m$ for $m_\chi=1$ TeV.}
    \label{fig:sigma}
\end{figure}
The value of the cross-section is normalized to produce one event in the window $215-270 $ keV nuclear recoil energy shown by the olive band.
The total event  rate in that window is given by,
\begin{eqnarray}
    R_{I}=~N_T\int_{215{\rm ~keV}}^{270{\rm ~keV}} \mathrm{d}E_R~ \frac{\mathrm{d} R}{\mathrm{d} E_R},
\end{eqnarray}
where $N_T=2.8\ \mathrm{tonne\text{-}year}$ for LZ~\cite{LZ:2026axp}.
As mentioned earlier, to address the observed event, we require
$R_{I}=1$.
However, from the same figure one can apprehend that, for smaller values of $\delta$, DM will produce event spectra in low $E_R$ as well.
To prevent any excess events in low recoil  energies, we also impose the cut off,
\begin{eqnarray}
    R_{II}=~N_T\int_{5.4{\rm ~keV}}^{215{\rm ~keV}} \mathrm{d}E_R~ \frac{\mathrm{d} R}{\mathrm{d} E_R}\lesssim 10^{-3} R_{\rm BKG}.
\end{eqnarray}
Here, $R_{\rm BKG}=1713$ represents the SM background in the low recoil energy range \cite{LZ:2026axp}.

In Fig.\ref{fig:sigma}(a) we display the variation of required $\sigma_{\rm SI}$  with $m_\chi$ for different values of $\delta m \in (360-690)$ keV.
The color map indicates the variation in $\delta m$.
With higher values of $\delta m$, the recoil spectra exceed the window, requiring a higher value of $\sigma_{\rm SI}$ or, in other words, a higher $g_{Z'}$ to compensate.
In Fig.\ref{fig:sigma}(b) we showcase the variation of the required $\sigma_{\rm SI}$  with $\delta m$ for a fixed $m_\chi=1$ TeV.
Note that the required cross-section to explain the LZ event is $\mathcal{O}(2)$ smaller in our model than in models with endothermic scattering.
Fig.\ref{fig:sigma}(b) also shows that the required parameter space is allowed by the constraints from solar capture and subsequent annihilation \cite{Pospelov:2026ewn,Bose:2026ndd, Nguyen:2026lui}.
In our model, there is no tree-level $Z'b\bar b$ or $Z' \nu \bar \nu$ coupling. The CKM-induced annihilation strength to $b$ channel or mixing-induced annihilation to $\tau$ pairs will also be suppressed by $\sim (g_{Z'}/100)^2$ and can be ignored.

\begin{figure}[!tbh]
    \centering
    \includegraphics[scale=0.6]{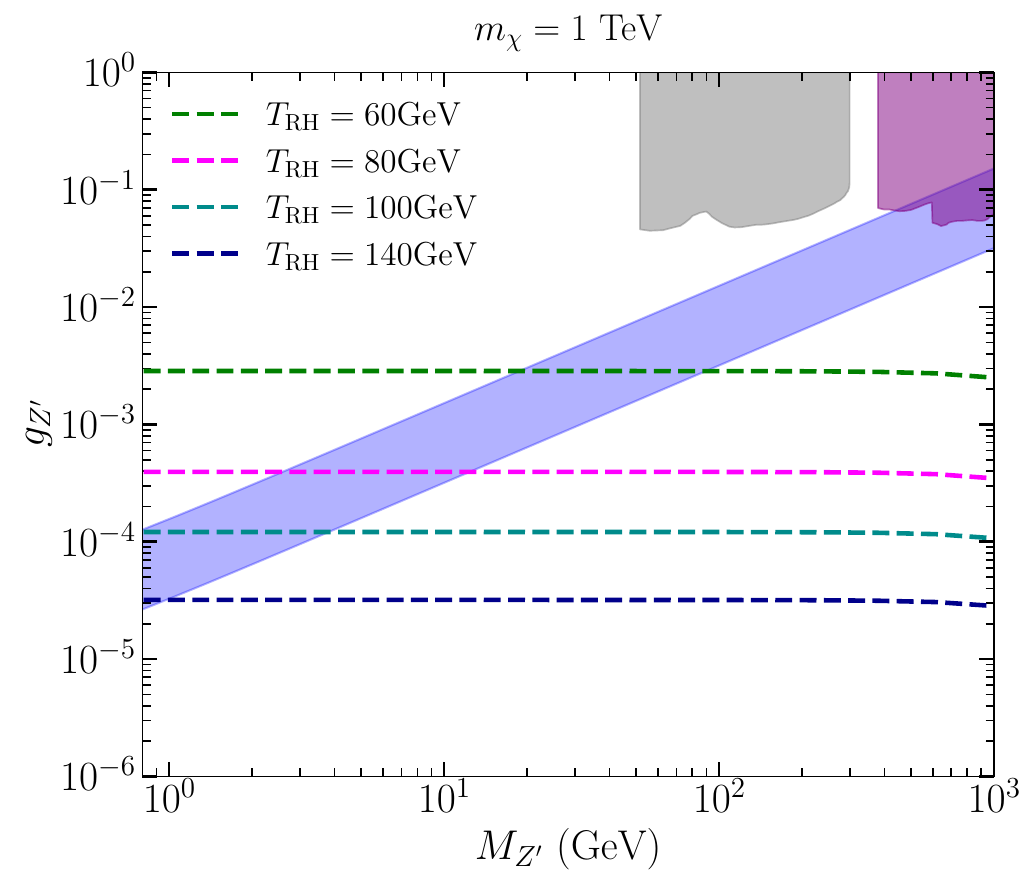}
    \caption{Required coupling with varying $Z'$ mass for exothermic scattering, shown by the blue shaded diagonal band for DM mass $1$ TeV.
    The dashed lines are the contours that satisfy the relic density with different values of $T_{\rm RH}$.}.
    \label{fig:relic}
\end{figure}
Finally, in Fig.\ref{fig:relic} we display the best-fit region from the LZ event in the $M_{Z'}$ vs. coupling parameter space for $m_\chi=1$ TeV indicated by the blue shaded diagonal band.
The parameter space that can satisfy the observed relic density for DM mass $1$ TeV with different values of $T_{\rm RH}\in (60,80,100,140)$ GeV is represented
by green, magenta, cyan, and blue
dashed lines, respectively.
We find that at low $T_{\rm RH}$, relic and LZ can both be explained simultaneously.
With a decrease in $T_{\rm RH}$, the required $g_Z'$ increases to compensate for the Boltzmann suppression.
However, in our chosen range with $m_{Z'}\ll 2 m_\chi$, the DM production rate as well as the DM abundance is mostly insensitive to $Z'$ mass.

So far we have not commented on the Yukawa sector involving the scalar doublets.
To generate the CKM matrix, we extend the model with 4 extra Higgs doublets. The Yukawa sector now reads as,
    \begin{equation}
        \begin{aligned}
            \mathcal{L}_{Y}=&-\sum_{i=1,2,3}\left(y^d_i \bar Q_{iL}H_0 d_{iR} + y^u_i\bar Q_{iL} \tilde{H}_0 u_{iR}\right)\\
            & -y^{d}_{12}\bar Q_{1L}d_{2R} H_4 - y^{d}_{13}\bar Q_{1L} d_{3R} H_2 - y^{d}_{21} \bar Q_{2L} d_{1R} H_3 \\
            &- y_{23}^d \bar Q_{2L} d_{3R} H_1 - y^d_{31} \bar Q_{3L}d_{1R} H_1 - y^d_{32} \bar Q_{3L}d_{2R} H_{2} \\
            & - y^{u}_{12} \bar Q_{1L} u_{2R} \tilde{H}_3 - y^u_{13}\bar Q_{1L}u_{3R}\tilde{H}_1- y^u_{21}\bar Q_{2L}u_{1R} \tilde{H}_4 \\
            & - y^{u}_{23}\bar Q_{2L}u_{3R} \tilde{H}_2 - y^u_{31} \bar Q_{3L}u_{1R} \tilde{H}_{2} - y^u_{32}\bar Q_{3L}u_{2R}\tilde{H}_1
          + h.c.
        \end{aligned}
    \end{equation}
After the symmetry breaking, we obtain the following forms of quark mass matrices:
    \begin{equation}
        M^{d} = \frac{1}{\sqrt{2}}
        \begin{pmatrix}
            y^d_1 v_0 & y^d_{12} v_4 & y^d_{13} v_2 \\
            y_{21}^d v_3 & y_2^d v_0 & y_{23}^d v_1 \\
            y_{31}^d v_1 & y_{32}^d v_2 & y_3^d v_0
        \end{pmatrix},
        \ \ \ \ 
        M^{u} = \frac{1}{\sqrt{2}}
        \begin{pmatrix}
            y^u_1 v_0 & y^u_{12} v_3 & y^u_{13} v_1 \\
            y_{21}^u v_4 & y_2^u v_0 & y_{23}^u v_2 \\
            y_{31}^u v_2 & y_{32}^u v_1 & y_3^u v_0
        \end{pmatrix}.
    \end{equation}
    
These mass matrices can be diagonalized by $U_{uL}^{\dagger} M_u U_{uR}
=
D_u
=
\operatorname{diag}(m_u,m_c,m_t),
$
and
$
U_{dL}^{\dagger} M_d U_{dR}
=
D_d
=
\operatorname{diag}(m_d,m_s,m_b).
$
The CKM matrix is then
$V_{\rm CKM}
=
U_{uL}^{\dagger} U_{dL}
$. 

However, this can attract strong constraints from meson mixings such as $K^{0}-\bar K^{0}$ mixing, $B_s-\bar B_s$ mixing, $B^{0}-\bar B^{0}$ mixing, $D^{0}-\bar D^{0}$ mixing due to the induced flavor-changing current. 
We first discuss $B_s-\bar B_s$ mixing as an example to show the possible problem that may arise from flavor mixing and how to solve it. For simplicity, we assume that $U_{uR}=U_{uL}=U_{dR}=I$ and thus we have 
$
        M_u =D_u,\ U^{\dagger}_{dL}M_d=D_d ,\  V_{\rm CKM}=U_{dL}.
$
And the mass eigen states $q^{m}_{L,R}$ are $d^{m}_R = d_R,\,d^{m}_L=V_{\rm CKM}^{\dagger}d_L,\,u^{m}_R=u_{R}^{},\,u^{m}_{L}=u_{L}^{}$.
    After rotating the quarks to their mass eigen states, we have 
    \begin{equation}
        \mathcal{L} \supset g_{Z'} Z^{'}_{\mu}\left[(-V_{ub}^{*}V_{us}+V_{cs}V_{cb}^{*})\bar b_{L}\gamma^{\mu}s_{L}+h.c.\right].
    \end{equation}
    The exchange of the $Z^{'}$ can generate the following four quark effective Lagrangian
    \begin{equation}
        \mathcal{L}_{B_s-\bar B_s} = -\frac{g_{Z'}^2(-V_{ub}^{*}V_{us}+V_{cs}V_{cb}^{*})^2}{2m_{Z^{'}}^2} (\bar b \gamma^{\mu}P_Ls)\bar b \gamma_{\mu}P_Ls + h.c.
    \end{equation}
    Using the matrix element~\cite{Lenz:2006hd}
        $\langle B_s|(\bar s \gamma^{\mu}P_Lb) (\bar s\gamma_{\mu}p_Lb)|\bar B_s\rangle=\frac{2}{3}m_{B_s}^2f_{B_s}^2B$
    with $f_{B_s}=230.3\ \rm MeV$~\cite{FlavourLatticeAveragingGroupFLAG:2024oxs} and $\ B = 0.813$~\cite{Dowdall:2019bea},
    we obtain the mixing mass parameter $\Delta m_s^{NP}$ as
    \begin{eqnarray}
        \Delta m_s^{NP} &=& -2 \frac{1}{2m_{B_s}}\langle B_s|\mathcal{L}_{B_s-\bar B_s}|\bar B_s\rangle = \left|\frac{g_{Z'}^2(-V_{ub}^{}V_{us}^{*}+V_{cs}^{*}V_{cb}^{})^2}{3m_Z^2}m_{B_s}f_{B_s}^2B\right|~~\nonumber\\
        &\approx& 1.9\left(\frac{g_{Z'}}{0.01}\right)^2\left(\frac{100\ \rm GeV}{m_{Z^{'}}}\right)^2\ \rm ps^{-1}.
    \end{eqnarray}
    This is too large compared with the experimental data \cite{ParticleDataGroup:2026mpi}.
    However, since there are enough free parameters, we can choose $U_{dR}=V_{CKM}^{'},\ U_{uR}=U_{uL}=I,\ U_{dL}=V_{\rm CKM}$ where $V_{CKM}^{'}$ is an unconstrained complex matrix and by appropriate choice of $V_{\rm CKM}^{'}$, this problem can be solved. For this scenario the mass eigen states become $u^{m}_{R}=u_R,\  u^{m}_{L}=u_L,\  d^{m}_{R}=V_{CKM}^{'\dagger}d_R,\  d^{m}_{L}= V_{CKM}^{\dagger}d_L$ and the matrix element reads as 
    \begin{equation}\label{eq:BBbar mixing}
        \begin{aligned}
            \langle B_s|-\mathcal{L}_{B_s-\bar B_s}|\bar B_s\rangle&
            =\frac{g_{Z'}^2 m_{B_s}^2}{3m_{Z^{'}}^2}\left[(A^2+A^{'2})f_{B_s}^2B- AA^{'}\left(\frac{m_{B_s}^2}{(m_s+m_b)^2}+\frac{3}{2}\right)f_{B_s}^2B_5\right]\\
            &=\frac{g_{Z'}^2 m_{B_s}^2f_{B_s}^2B A^2} {3m_{Z^{'}}^2}\left[k^2+1-\left(\frac{m_{B_s}^2}{(m_s+m_b)^2}+\frac{3}{2}\right)\frac{B_5}{B} k\right].
        \end{aligned}
    \end{equation}
  where $A=-V_{ub}^{*}V_{us}+V_{cs}V_{cb}^{*},\ A^{'}=-V_{ub}^{'*}V^{'}_{us}+V^{'}_{cs}V_{cb}^{'*}$ , $A^{'}=k A$ and $k$ is an arbitrary complex number.  We can choose an appropriate value of $k$ so that  eq.\eqref{eq:BBbar mixing}  equals zero and can evade the constraint from the $B_s-\bar B_s$ mixing.  For brevity, we do not explicitly discuss the other meson mixing issues and other rare decays which can be fixed by the similar cancellation mechanism exploiting the different free parameters in the mixing matrices as shown above.

The collider constraint on $Z'$ from CMS dijet searches excludes $g_{Z'}\gtrsim 0.1$ for $M_{Z'}\lesssim 140$ GeV with ISR \cite{CMS:2026aim} and ATLAS dijet search excludes $g_{Z'}\gtrsim 0.04$ for $M_{Z'}\gtrsim 400$ GeV \cite{ATLAS:2025okg} depicted by the magenta shaded region in Fig.\ref{fig:relic}.
Also, the CMS searches for $Z'$ decay to $q\bar{q}$ constrain the $g_{Z'}$ in the heavier $M_{Z'}$ range \cite{CMS:2026yvw} shown by the gray shaded region in Fig.\ref{fig:relic}.
Mixing-induced di-lepton constraints from LHC \cite{Amrith:2018yfb} are also weaker for our scenario.
Hence, the DM framework analyzed in this $q_1-q_2$ setup is also allowed from the existing constraints on $Z'$.
    
{\it Conclusion.} 
Indeed, the recent observation of the isolated event by LZ \cite{LZ:2026axp} has already stimulated efforts by DM model builders to look for a possible DM interpretation and has resurrected the heavy DM paradigm.
While most of the literature discusses endothermic scattering by a heavy DM, such as Higgsino and other flavor-universal gauged $U(1)_X$ DM models, they are in stringent tension with the constraint from possible annihilation signals to $\nu\bar \nu$ or $b\bar b$.
In this paper, we consider a hadrophilic fermion inelastic DM coupled to a $q_1-q_2$ gauge boson to address this LZ event.
We consider the TeV-scale DM to be produced at a low $T_{\rm RH}$ with a coupling strong enough to produce such a recoil.
However, the required coupling/ interaction rate is lower than the typical thermal DM paradigm.
Both $\chi_1$ and $\chi_2$ are produced in equal fractions with a tiny mass splitting $\delta m\sim \mathcal{O}(100)$ keV and both survive at the cosmological scale.
Such a scenario is difficult to realize in the thermal DM paradigm.
The absence of any tree-level coupling to $b$ quark or neutrino makes the DM allowed from the indirect searches.
However, such flavor structures induce some flavor anomalies, which can also be resolved with proper rotation in the quark mass matrices.
We show that the parameter space where a TeV-scale DM produces isolated event is in good agreement with the relic density and is also allowed from the existing constraints on the BSM gauge boson. 
This will motivate future studies with flavor-dependent hadrophilic DM production invoking low reheating temperature without the prior assumption of instantaneous reheating. Rigorous analysis of collider signature prospects of the BSM particles will also bring novel phenomenological insights.

{\bf Note added.} Ref.\cite{Lee:2026wof} with similar model  appeared at the final stage of this work.
However, ref.\cite{Lee:2026wof} only considered $B_1-B_3$ model with a thermal freeze-out scenario.
On the other hand, we consider freeze-in production at a low reheating temperature and exothermic scattering to produce the recoil event realized in a different parameter space.

    
\section*{Acknowledgments}
SJ thanks Anjan K. Barik and Anirban Majumdar for helpful discussions.
 X.-G.H. was supported by the Fundamental Research Funds for the
Central Universities, by the National Natural Science Foundation of the People’s Republic
of China (12090064, 12375088, and W2441004). The work of SJ is funded by the National Natural Science Foundation of China (12425506 and 12375101) and the
State Key Laboratory of Dark Matter Physics.

	\bibliographystyle{JHEP} 
	\bibliography{refs.bib}

\end{document}